\documentclass[12pt]{iopart}
\usepackage{graphicx}
\usepackage{color}
\usepackage{csquotes}
\usepackage{amssymb}
\usepackage{multirow}

\newcommand{\ket}[1]{|#1 \rangle}

\newcommand{\pra}{\textit{Phys. Rev.} A }

\begin{document}

\title{Adiabatic two-photon quantum gate operations using a long-range photonic bus}%

\author{Anthony P Hope$^{1}$, Thach G Nguyen$^{1}$, Arnan Mitchell$^{1}$ and Andrew D Greentree$^{2}$}%

\address{$^1$ ARC Centre of Excellence for Ultrahigh bandwidth Devices for Optical systems (CUDOS) and School of Electrical and Computer Engineering, RMIT University, Melbourne, Australia}
\address{$^2$ Chemical and Quantum Physics, School of Applied Sciences, RMIT University, Melbourne, Australia}
\eads{anthony.hope@rmit.edu.au}

\begin{abstract}
Adiabatic techniques have much potential to realise practical and robust optical waveguide devices.  Traditionally photonic elements are limited to coupling schemes that rely on proximity to nearest neighbour elements. We combine adiabatic passage with a continuum based long-range optical bus to break free from such topological restraints and thereby outline a new approach to photonic quantum gate design.  We explicitly show designs for adiabatic quantum gates that produce a Hadamard,  50:50 and 1/3:2/3 beam splitter, and non-deterministic CNOT gate based on planar thin, shallow ridge waveguides.  Our calculations are performed under conditions of one and two-photon inputs.

\end{abstract}

\pacs{05.60.Gg, 42.82.Ds, 42.82.Et, 03.67.-a}

\maketitle

\section{Introduction}
Integrated optics is becoming one of the most important platforms for the production of compact, scalable, linear optical quantum devices \cite{bib:OFV2009}.  Much of this progress derives from the use of laser-defined waveguides in glass or polymer, which enables compact three-dimensional waveguide geometries to be designed and rapidly prototyped \cite{bib:MDG+2012}.  Whilst these devices are clearly important for scientific applications, they are not compatible with standard lithographic processes and are limited in the topologies that can be considered.  Direct write waveguides rely on serial material modification which places limitations on the complexity of the waveguide system designs that can be achieved.  As a result, most geometries rely on evanescent coupling of nearest neighbours.  It has recently been shown that long range coupling can be achieved through utilisation of lateral leakage radiation in thin, shallow ridge silicon photonic waveguides \cite{bib:AH2013}.  This paper shows that the ability to break free from the limitations of simply connected waveguide topologies offers new opportunities for the realisation of complex, multi-port quantum gates.

One of the most fundamental elements required for integrated optical devices, especially quantum devices, is the beamsplitter.  This is the essential element for any interferometer, and can also be used (with trivial phase control) to effect a Hadamard rotation \cite{bib:CAK1998}.  In the two-photon subspace, the beamsplitter can show the Hong-Ou-Mandel effect \cite{bib:HOM1987}, one of the clearest non-trivial experiments to highlight the fundamental differences between classical and quantum optics.  In integrated optics, a beamsplitter is typically realised through the use of a directional coupler.  This is a device where two waveguides are brought into close proximity so that evanescent coupling causes population to tunnel between the waveguides.  Truncating the device to the appropriate length then effects the desired beam splitting ratio.  Although, in principle, it should be relatively easy to build such devices; in practice, any lack of control in the actual waveguide size leads to a lack of control in the evanescent tunnelling, and hence the length of device required to achieve a particular beam splitting ratio will be effectively unknown.  A common solution is to post-select devices from a suite of similar devices, or alternatively, phase shifting elements such as heaters, can be used to fine tune and reconfigure devices \cite{bib:SVP2012}.

Adiabatic passage promises a solution to issues of device variability that require post-fabrication tuning.  This is because adiabatic evolution goes as the ratio of tunnel matrix elements, rather than the absolute value of those elements.  The tradeoff is that adiabatic devices are typically longer than their non-adiabatic counterparts, and whether the rewards of seeking an adiabatic vs non-adiabatic device are justified depends on the degree of device control required and the available footprint.  Adiabatic methods for transport of population between states of the kind we envisage here began with the STIRAP (STImulated Raman Adiabatic Passage) protocol \cite{bib:GRB+1988,bib:KTS2007}, where robust transfer of population between atomic energy levels is effected by laser control, and the all-spatial variant that is sometimes called CTAP (Coherent Tunnelling Adiabatic Passage) \cite{bib:ELC+2004,bib:GCH+2004,bib:GDH2006,bib:Pas2006,bib:LDO+2007}.

\begin{figure}[tb!]
\centering
\includegraphics[width=0.6\textwidth]{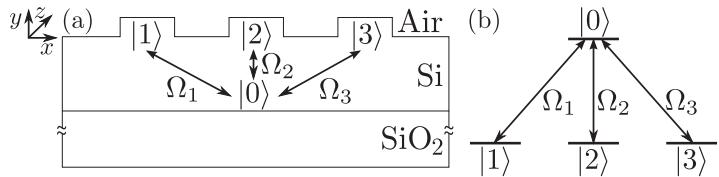}
\caption{(a) Schematic showing three thin shallow ridge SOI waveguides with a common bus (slab) mode.  Coupling between waveguide and slab is effected by the overlap integral between the waveguide and the slab, and is therefore a function of the absolute position of the waveguide with respect to the slab.  Each waveguide is assumed to be isolated from the other waveguides so that there is no appreciable evanescent tunnelling between the waveguides.  (b) The tripod atom is the simplest system to realise USB style geometric adiabatic gates.  Three ground states, $\ket{1}$, $\ket{2}$, and $\ket{3}$, are coupled to a single excited state $\ket{0}$ via optical fields with Rabi frequencies $\Omega_i$ for the transition between $\ket{i}$ and $\ket{0}$. The waveguide modes are equivalent to the ground states of the tripod atom, whilst the shared bus mode plays the role of the excited state.}
\label{fig:tripod}
\end{figure}

Here we explore theoretically the potential for effecting geometric gates via long-range CTAP in thin, shallow ridge silicon-on-insulator (SOI) waveguides, shown schematically in figure~\ref{fig:tripod} (a).  Specifically, we demonstrate several important gate designs including 50:50 beamsplitter and 1/3:2/3 beamsplitter, using a spatial version of the method outlined by Unanyan, Shore and Bergmann (USB) \cite{bib:USB1999}.  Our calculations are performed for both one and two photon input states.  Further, we concatenate these devices to show an adiabatic non-deterministic CNOT gate, following the approach described by Ralph \textit{et al.} \cite{bib:RLB2002}.  Our scheme utilises a long-range common bus mode present in thin, shallow ridge waveguide devices \cite{bib:AH2013}.  This common bus mode provides a significant, new opportunity to develop planar geometries which are nonetheless not restricted to linear nearest-neighbour coupling.  In this way, we see our approach as being more amenable to mass production, especially CMOS compatible fabrication, than truly three-dimensional approaches such as those described in, for example, references~\cite{bib:MDG+2012,bib:HGH2011,bib:RV2012}.

This paper is organised as follows.  We first provide a brief introduction to thin, shallow ridge SOI waveguides, with emphasis on their effective refractive index and coupling to slab modes.  As will be shown, control of both the magnitude \emph{and} the sign of the coupling between waveguides and the slab can be achieved by the position of the waveguide relative to the slab mode.  A change in the sign of the coupling leads to symmetry breaking mechanisms that are essential for USB-style geometric gates.  With this understanding, we generate an effective Hamiltonian that can be used to effect arbitrary geometric gate sequences, and in particular we describe methodologies to realise an adiabatic power splitter, a Hadamard gate and 1/3:2/3 beamsplitter via the USB approach.  After demonstrating one-qubit gates, we show the extension to two-photon gates, in particular showing that the well-known Hong-Ou-Mandel effect is preserved under conditions of adiabatic passage.  Finally, we show the full state evolution for a non-deterministic linear optical controlled gate operating in the coincidence basis.  This non-trivial two-photon, two-qubit entangling gate is completely simulated across eight optical modes (seven waveguide modes plus one bus mode), each of which can potentially have either 0, 1 or 2 photons.

\section{Adiabatic evolution with thin, shallow ridge silicon-on-insulator waveguides}

Thin, shallow ridge SOI waveguides when operating in TM polarisation can exhibit lateral leakage behaviour \cite{bib:WPM2007}.  Photons in a TM guided mode can leak into a lateral unguided TE slab mode, propagating at a specific angle to the waveguide axis due to polarisation conversion at the ridge side-walls \cite{bib:Thach}.  This TE slab mode can act as a bus mode to allow long-range communication between isolated waveguides \cite{bib:AH2013,bib:Naser-OE}.

When the silicon slab is terminated, the continuum of TE slab radiation is discretised into discrete TE slab modes.  On careful selection of the slab width, one of these TE slab modes can be phase-matched to the TM guided mode.  Thus, the photons from the guided TM mode can couple to this TE slab mode.  One particular method for varying the strength of the coupling between the guided TM mode and TE slab mode is by varying the relative location of the ridge waveguide on the slab \cite{bib:AH2013,bib:Naser-OE}.  This technique opens up a new class of coupler that can enable interactions between multiple, well separated waveguides simultaneously, which is not possible in a traditional planar evanescent arrangement and has recently been proposed for CTAP devices \cite{bib:AH2013}.  
Whilst here we only consider interactions mediated via a single, discrete slab mode for clarity, generalising our method so that coupling is via continuum states should be possible following the methods in references \cite{bib:PH2007,bib:DSH2009}.

Considering a discrete silicon slab supporting a laterally defined bus mode $\ket{0}$ with propagation constant $\beta_{0}$ and $N$ forward propagating waveguide modes $\ket{i}$ of $\beta_{i}=k_0n_i$ with effective index $n_i$, where $k_0=2\pi/\lambda_{0} $ is the free space wavevector for wavelength $\lambda_{0}$.  Under these definitions, we may write down the system using a tight-binding Hamiltonian in second quantised form as:

\begin{eqnarray}
H(z) = \beta_0 a^{\dag}_0 a_0 + \sum_{i = 1}^N \beta_i a^{\dag}_{i}a_{i} + \Omega_i a^{\dag}_{0}a_{i} + h.c.  \label{eq:Ham}
\end{eqnarray}
where $a_i$ $(a^{\dag}_i)$ is the photon annihilation (creation) operator acting on mode $\ket{i}$ for $i = 0...N$.  Each waveguide is mutually isolated by separating an appropriate distance, ensuring there is no appreciable evanescent coupling, so that the waveguides only communicate through the common bus.  The strength of this coupling ($\Omega_{i}$) is controlled by translating the waveguides laterally across the slab and this response is sinusoidal due to the nature of the bus mode \cite{bib:AH2013,bib:Naser-OE}.  The coupling of a single waveguide to the bus is $\Omega(z) = \Omega_{\max} \sin\lbrack\beta_{0} x(z)\rbrack$, where $x$ is the lateral waveguide location, assuming isolation occurs in the centre of the slab.  The lateral waveguide dimension in turn varies as a function of the propagation dimension, $z$, which is the mechanism to effect the adiabatic passage.  The relationship between the lateral ($x$) and forward propagation ($z$) dimensions is  controlled so that the couplings are varied adiabatically.  The maximum coupling ($\Omega_{\max}$) available depends on the waveguide dimensions \cite{bib:Thach,bib:Ravi} and can be calculated as the overlap integral between the bare bus TE and waveguide TM modes.  This magnitude is represented as the imaginary effective index of a TM-TE coupled mode on an open slab \cite{bib:Thach}, or in the discrete case by observing the level of mode splitting throughout translation \cite{bib:AH2013}.  In the case where the slab width increases to accommodate additional waveguides, it is expected that this coupling will decrease as the maximal overlap of the single bus mode at any point decreases, resulting in longer devices.  This scalability is of importance when considering more general Morris-Shore type devices \cite{bib:MS1983,bib:RVS2006}.

In the discussion that follows we will adopt two separate notations.  When we consider only the one-photon subspace, we will use the compact notation of defining the basis states by the position of the photon, i.e. we define $\ket{i} \equiv a_i^{\dag}\ket{\emptyset}$ where $\ket{\emptyset}$ is the vacuum state of the system.  However, when we deal with two-photon states, we will define the states by the occupation numbers of each mode, so for example the state $\ket{0110} \equiv a_1^{\dag} a_2^{\dag} \ket{\emptyset}$. All of our simulations use a tight binding approach to solving the spatially varying Hamiltonian, and do not assume that the adiabatic limit is achieved.

\section{Adiabatic power divider}

Adiabatic techniques can be used for power division applications either by fractional adiabatic passage \cite{bib:DOS+2009,bib:MLC+2012} or the use of additional waveguide modes \cite{bib:CKR+2012,bib:CCR+2012}.  The presence of a shared bus offers an intriguing alternative technique.  We first consider three waveguides acting as a tripod atom connected via a bus state (figure \ref{fig:tripod}), the bus can be designed to be degenerate with the individual isolated waveguides, which will result in improved adiabaticity.  All of our modelling was performed using the tight-binding Hamiltonian of equation~\ref{eq:Ham} and was conducted in the adiabatic limit.

\begin{figure}[tb!]
\centering
\includegraphics[width=\textwidth]{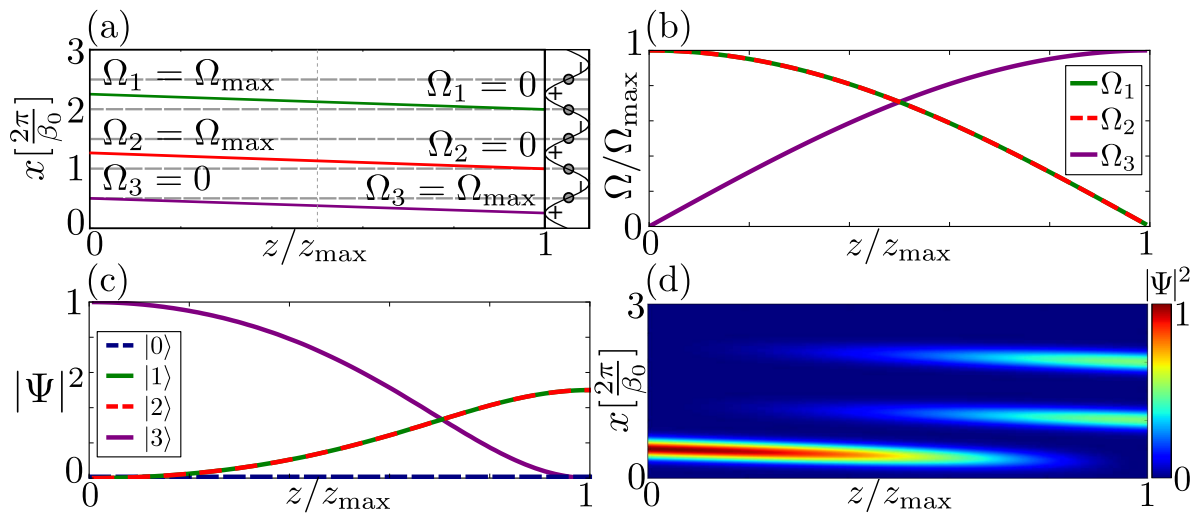}
\caption{Three waveguide power divider taking population from $\ket{3} \to \ket{1}+\ket{2}$, where the parameters for each waveguide are coloured consistently as green $(\ket{1})$, red $(\ket{2})$ and purple $(\ket{3})$, the population of the bus mode ($\ket{0}$) is blue;
(a) Positioning the waveguide centres to provide counter-intuitive multiple waveguide population transfer, the sidebar to the right shows the magnitude of the sinusoidal coupling between the waveguides and the slab mode, based on the modal overlap.
(b) Magnitude of coupling terms using the provided waveguide geometries.
(c) Modal population of all states throughout system evolution shows 50:50 power division to output waveguides. (d) Power as a function of position through the device, observe no measurable population in the slab at any time through the protocol. 
}
\label{fig:ps}
\end{figure}

Here we wish to inject light into port $\ket{3}$ and arrive in an even superposition of both $\ket{1}$ and $\ket{2}$.  The position of each waveguide is selected to provide ideal initial CTAP conditions [$\Omega_{3}(0)=0, \Omega_{1}(0)=\Omega_{2}(0)=\Omega_{\max}$].  
By translating the waveguides linearly across the slab, as illustrated in figure \ref{fig:ps}(a), the couplings are varied sinusoidally to effect the counter-intuitive pulse sequence.  In particular, we have $\Omega_3(z) = \Omega_{\max} \sin \left[ \pi z/(2 z_{\max})\right]$ and $\Omega_1(z) = \Omega_2(z) = \Omega_{\max} \cos \left[ \pi z/(2 z_{\max})\right]$ shown in figure \ref{fig:ps}(b).  The $\sin/\cos$ coupling scheme has the nice property that the adiabaticity is constant throughout the protocol \cite{bib:CH1990,bib:LS1996,bib:VG2013}.  As $\Omega_{1}$ and $\Omega_{2}$ remain identical, an even power split arrives in each waveguide with the populations throughout the protocol described in figure \ref{fig:ps}(c).  Combining the predicted paths, calculated population values and expected Gaussian mode profile of the waveguides gives a more visual representation of this transfer (figure \ref{fig:ps}(d)).

This one input-two output device is equivalent to a Y-splitter but without the conventional restrictions of close proximity or ordering of the waveguides.  There is therefore enhanced flexibility with the bus approach than more conventional approaches.  This technique can also be extended to distribute population evenly across many waveguides in a method akin to that in reference \cite{bib:RV2012}.  The overall device length required to perform a successful adiabatic passage depends on $\Omega_{\max}$.  The waveguide dimensions specified in \cite{bib:AH2013} provided a coupling length of 150 $\mu$m for a single waveguide and bus.  
It was shown that device lengths of $z_{max} \ge 2 $mm were required to successfully achieve robust adiabatic transport between two waveguides using the long-range bus, without significant population of the bus.  This implies that to achieve successful CTAP behaviour the total device length must be at least longer than 15 coupling lengths.  
The magnitude of maximum coupling $\Omega_{\max}$ can be enhanced through waveguide engineering \cite{bib:Ravi,bib:KKS2008}.  Increasing the available coupling will decrease the coupling length and hence reduce the absolute device length $z_{\max}$ required to maintain adiabaticity, although such optimisation is not critical to explain our concepts.

\section{Controlled-ratio beamsplitters}\label{sect:CRBS}
The method of power splitting can be modified to effect robust quantum gates via the USB method \cite{bib:USB1999}.  In this process, a double application of the power division is applied, with a change in the sign of the coupling applied between the first and second applications of the splitting.  Because of the standing wave nature of the bus mode, the coupling between the waveguide and the slab varies sinusoidally with the waveguide position. The sinusoidal variation means that the sign of the coupling reverses in each standing-wave period.  By ensuring that the forward and backward adiabatic passage crosses periods with the appropriate signs,  the necessary symmetry breaking that is at the heart of the USB process can be achieved by waveguide translation alone.

In the one-photon subspace, the quantum gate is specified without loss of generality to act on qubit subspace of $\ket{1}$ and $\ket{2}$, with $\ket{3}$ as an auxiliary mode, explained below. We assume that the system is initialised in an arbitrary superposition $\ket{\psi}= \gamma_{1} \ket{1} + \gamma_{2} \ket{2}$, and for convenience we express the qubit in the dark/bright state basis,

\begin{eqnarray}
\ket{D} = \frac{\Omega_{2} \ket{1} - \Omega_{1} \ket{2}}{\sqrt{\Omega_{1}^2 + \Omega_{2}^2}}, \quad \ket{B} = \frac{\Omega_{1} \ket{1} + \Omega_{2} \ket{2}}{\sqrt{\Omega_{1}^2 + \Omega_{2}^2}}.
\end{eqnarray}
Note that provided the ratio of the couplings remains constant, the compositions of the dark and bright states will not change.  Ideally $\ket{D}$ remains completely isolated from $\ket{0}$, whilst $\ket{B}$ can be influenced using CTAP.  The counter-intuitive sequence transfers $\ket{B} \to \ket{3}$, instigating a phase reversal of $\ket{B}$ before returning it will alter the superposition, performing a rotation of the single qubit.  
The magnitude of this rotation is set by $\alpha=\Omega_{2}/\Omega_{1}$, and the net effect of a double application of the adiabatic passage is the gate:
\begin{eqnarray}
G = \frac{1}{1 + \alpha^2}\left[\begin{array}{cc} \alpha^2 - 1 & -2\alpha \\ -2\alpha & 1-\alpha^2 \end{array}\right].
\label{eq:G}
\end{eqnarray}
We illustrate the operation of this gate with three examples.

\begin{figure}[tb!]
\centering
\includegraphics[width=\textwidth]{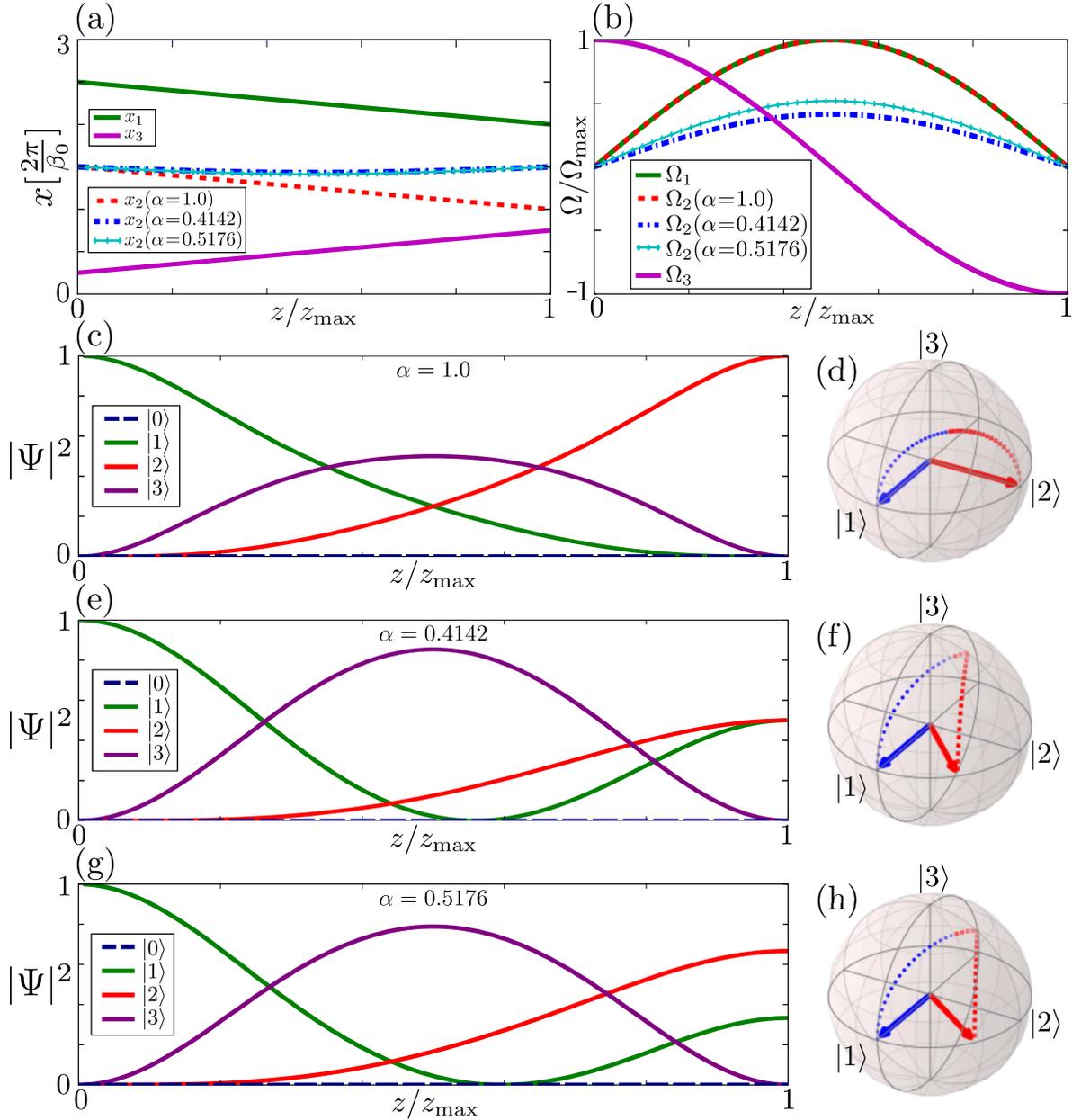}
\caption{Single photon arbitrary X-Z gate operations;
(a) the paths of $\ket{1}$ and $\ket{3}$ are unchanged in each case, where the different gate operations are provided by altering the trajectory of $\ket{2}$ which is represented as red (dashed), blue (dot-dashed) and cyan (plus markers) for $\alpha=$1, 0.4142 and 0.5176 respectively,
(b) the coupling of $\ket{3}$ is now allowed to turn negative instigating a break in the symmetry of forward and backwards paths,
(c) $\alpha=1$ operates as a NOT gate completely transferring population from $\ket{1} \to \ket{2}$,
(d) qutrit representation showing the forward (blue) and backwards (red) dotted paths taken through evolution,
(e)-(f) $\alpha=0.4142$ conforms to a Hadamard operation creating a 50:50 superposition of $\ket{1}$ and $\ket{2}$,
(g)-(h) $\alpha=0.5176$ creates a 1/3:2/3 beamsplitter.
}
\label{fig:3-phase}
\end{figure}

Preparing the system with all population initially in $\ket{1}$, the waveguides are translated to provide counter-intuitive transfer to $\ket{3}$, $\Omega_{3}$ then goes negative transferring back to $\ket{1}$ and $\ket{2}$ (figure \ref{fig:3-phase}(a)-(b)).  Translating all waveguides linearly ($\alpha=1$) results in complete population transfer to $\ket{2}$ (figure \ref{fig:3-phase}(c)).  Performing this operation takes the system from $\ket{\psi_0}=\gamma_{1} \ket{1} + \gamma_{2} \ket{2} \to \ket{\psi_{f}}=\gamma_{2} \ket{1} + \gamma_{1} \ket{2}$, which is an X-gate: the quantum equivalent of a NOT gate.  The trajectory taken by the population is also depicted on the sphere shown in figure~\ref{fig:3-phase}(d). This qutrit representation displays only the real part of the state amplitudes \cite{bib:AG2002,bib:AK1996,bib:JAV1998}

Other gate operations occur when $\alpha \neq 1$.  The available gates are confined to rotations in the canonical X-Z plane for the qubit defined across modes $\ket{1}$ and $\ket{2}$ .  If we define the trajectory of the state with greatest coupling to be linear, i.e. so that the coupling between this waveguide and the slab is sinusoidal and maximal, then it follows that the other must trace out a curved trajectory.  So without loss of generality, assuming $\Omega_1 > \Omega_2$, we have $\Omega_{1}=\Omega_{\max}\sin(\beta_{0} x_{1}), \Omega_2 = \Omega_{\max}\sin(\beta_0 x_2) = \alpha \Omega_1, \therefore x_{2}=\sin^{-1}(\alpha \Omega_{1}/\Omega_{\max})/\beta_{0}$.

The Hadamard gate is a commonly used quantum information primitive, and is equivalent (up to phase) to a conventional beamsplitter.
Preparing the system entirely in $\ket{1}$ a successful Hadamard operation will result in the state being transformed to $(1/\sqrt{2})(\ket{1}+\ket{2})$.  Using (\ref{eq:G}), a value of $\alpha=\tan(\pi/8)\approx 0.4142$ provides this behaviour.  The waveguide trajectories required to provide this value of $\alpha$ are shown in figure \ref{fig:3-phase}(a), with the resulting coupling scheme in figure \ref{fig:3-phase}(b).  Evolving this system adiabatically results in the expected Hadamard operation as demonstrated in figure \ref{fig:3-phase}(e)-(f).

Another important beamsplitter is the 1/3:2/3 beamsplitter.  Beamsplitters with this splitting ratio form the basis for non-deterministic linear optical quantum computing \cite{bib:RLB2002,bib:HT2002}.  A suitable two port 1/3:2/3 beam splitter is designed with $\alpha=0.5176$ and operation is shown in figure \ref{fig:3-phase}(g)-(h).

\section{Two-photon operation}

There are few studies that have explicitly considered adiabatic passage of more than one particle \cite{bib:BRG+2012,bib:Ols2014,bib:Lon2014}, without invoking some mean-field or other effective treatment (as in for example references.~\cite{bib:GKW2006,bib:RCP+2008}).  We are not aware of any previous works that consider adiabatic multi-particle gates such as we describe here, and hence some explanation of the two-photon gate operation is required.

The one-photon calculations described above are indistinguishable from the results that would be obtained for a purely classical modal analysis.  Although the operation of the adiabatic gates on two-photon states is exactly what should be predicted from an equivalent conventional device, the microscopic details of how the adiabatic device achieves two-photon interactions is interesting and non-trivial.  

\begin{figure}[tb!]
\centering
\includegraphics[width=0.7\textwidth]{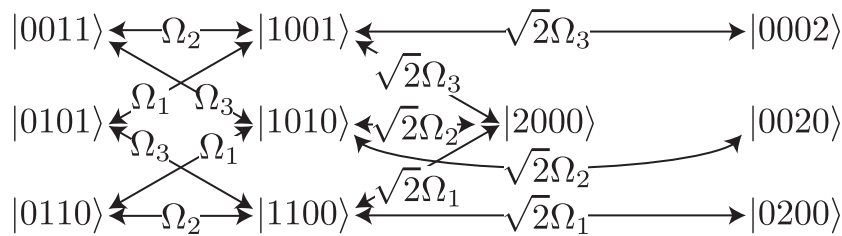}
\caption{There are ten states involved in the two-photon, four mode Hamiltonian.  This figure indicates the pertinent states and the strengths of the couplings between them
}
\label{fig:TenStateDiag}
\end{figure}

For two photons across four modes (three waveguides + one bus mode), there are ten states that need to be considered.  These, along with the couplings between the modes, are shown in figure~\ref{fig:TenStateDiag}.  The states are: $\ket{0011}$, $\ket{0101}$, $\ket{0110}$, $\ket{1001}$, $\ket{1010}$, $\ket{1100}$, $\ket{0002}$, $\ket{0020}$, $\ket{0200}$, and $\ket{2000}$, where as before the most significant digit denotes the number of photons in the bus mode, and the subsequent digits refer to the number of photons in waveguides one to three.  The Hamiltonian of the two-photon states is spanned by a four-dimensional null space comprising the (unnormalised) vectors
\begin{eqnarray}
\ket{D_1^{(2)}} &=& -\frac{-\Omega_1^2 + \Omega_2^2 + \Omega_3^2}{\sqrt{2} \Omega_2 \Omega_3}\ket{0011} -\frac{\Omega_1^2 - \Omega_2^2 + \Omega_3^2}{\sqrt{2} \Omega_1 \Omega_3}\ket{0101}  \nonumber \\
& & -\frac{\Omega_1^2 + \Omega_2^2 - \Omega_3^2}{\sqrt{2} \Omega_1 \Omega_2}\ket{0110} + \ket{2000}, \\
\ket{D_2^{(2)}} &=& \frac{\Omega_1^2}{\sqrt{2} \Omega_2 \Omega_3}\ket{0011} -\frac{\Omega_1}{\sqrt{2} \Omega_3}\ket{0101} -\frac{\Omega_1}{\sqrt{2} \Omega_2}\ket{0110} + \ket{0200}, \\
\ket{D_3^{(2)}} &=& -\frac{\Omega_2}{\sqrt{2}  \Omega_3}\ket{0011} +\frac{\Omega_2^2}{\sqrt{2} \Omega_1\Omega_3}\ket{0101} -\frac{\Omega_2}{\sqrt{2} \Omega_1}\ket{0110} + \ket{0020}, \\
\ket{D_4^{(2)}} &=& -\frac{\Omega_3}{\sqrt{2} \Omega_2}\ket{0011} -\frac{\Omega_3}{\sqrt{2} \Omega_1}\ket{0101} +\frac{\Omega_3^2}{\sqrt{2} \Omega_1\Omega_2}\ket{0110} + \ket{0002}.
\end{eqnarray}

\begin{figure}[tb!]
\centering
\includegraphics[width=0.7\textwidth]{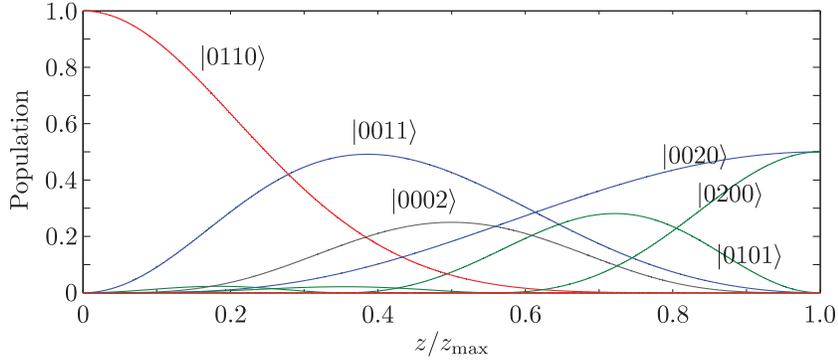}
\caption{State evolution during the Hong-Ou-Mandel type process (two photons initially in modes 1 and 2).  Population is initially in $\ket{0110}$, adiabatically transferred to a superposition of $\ket{0110}$, $\ket{0011}$, $\ket{0101}$, $\ket{0002}$, $\ket{0020}$ and $\ket{0200}$ at the midpoint.  Then the sign of the coupling on $\Omega_3$ is reversed, and the resulting interference leads to the entangled state $(1/\sqrt{2})(\ket{0200} - \ket{0020})$, as expected for a conventional beamsplitter interaction with two indistinguishable photons.
}
\label{fig:HOMAP}
\end{figure}

With this four-dimensional null space, it is difficult to gain insight into the exact properties of the null state for any given problem, but there are some important features that can be gleaned.  Firstly, notice that there is no overlap with states with a single photon in the bus mode.  This is desirable as it minimises sensitivity to loss from this mode.  There is, however, potential overlap with the bus mode from $\ket{D_1^{(2)}}$.  We have numerically confirmed that providing the system is initialised in either $\ket{0011}$, $\ket{0101}$ or $\ket{0110}$, there is no population in $\ket{2000}$ (up to the limits of the adiabaticity of our numerical experiment, as seen for example in figure~\ref{fig:HOMAP}), indicating that $\ket{D_1^{(2)}}$ is \emph{not} populated during the adiabatic gate operation.  The absence of population in the bus is important as it means that the adiabatic gate is indeed robust against loss from the bus mode.

Considering the 50:50 beamsplitter for the case of two indistinguishable input photons.  The system is prepared in the state $a_1^{\dag}a_2^{\dag}\ket{\emptyset} = \ket{0110}$.  Using the same coupling scheme as in the one-photon case, namely
\begin{eqnarray}
\Omega_1 &=& \Omega \sin\left(\pi z/z_{\max}\right), \\
\Omega_2 &=& \Omega \tan(\pi/8) \sin\left(\pi z/z_{\max}\right), \\
\Omega_3 &=&\Omega \cos\left(\pi z/z_{\max}\right), 
\end{eqnarray}
our results are shown in figure~\ref{fig:HOMAP}.  Note the smooth, adiabatic evolution.  In this case, the initial state $\ket{0110}$ is transformed to an entangled state at the midpoint of the protocol, with non zero population in $\ket{0110}$, $\ket{0011}$, $\ket{0101}$, $\ket{0002}$, $\ket{0020}$ and a very small contribution from $\ket{0200}$. This evolution should be contrasted with the corresponding case from the one-photon Hadamard interaction [figure~\ref{fig:3-phase}(e), (f)], where the bright state was entirely transported to the state $\ket{0001}$.  At the midpoint, there is a sign change on $\Omega_3$.  This sign change leads to interference, with the net result that at the end of the protocol, the entangled state $(1/\sqrt{2})(\ket{0200} - \ket{0020})$.  This state is precisely the state that provides the expected two-photon Hong-Ou-Mandel response as measurement of the output ports will project the photons to be either both at waveguide 1, or both at waveguide 2.

\section{CNOT operation}

The Controlled Not (CNOT) gate is a fundamental entangling gate and popular choice as member of a universal gate set for quantum computing \cite{bib:BBC+1995}.  This is a two qubit gate, where the state of the target is flipped conditional on the state of the control qubit.  One method to generate a scalable, but non-deterministic CNOT gate between individual photons is through combinations of linear optical elements (beamsplitters) \cite{bib:KLM2001}.  Here we show the set of results when applying one particular implementation (the coincidence basis implementation) of a non-deterministic CNOT gate, based on 1/3:2/3 and 50:50 beamsplitters, following Ralph \textit{et al.} \cite{bib:RLB2002}.

\begin{figure}[tb!]
\centering
\includegraphics[width=0.8\textwidth]{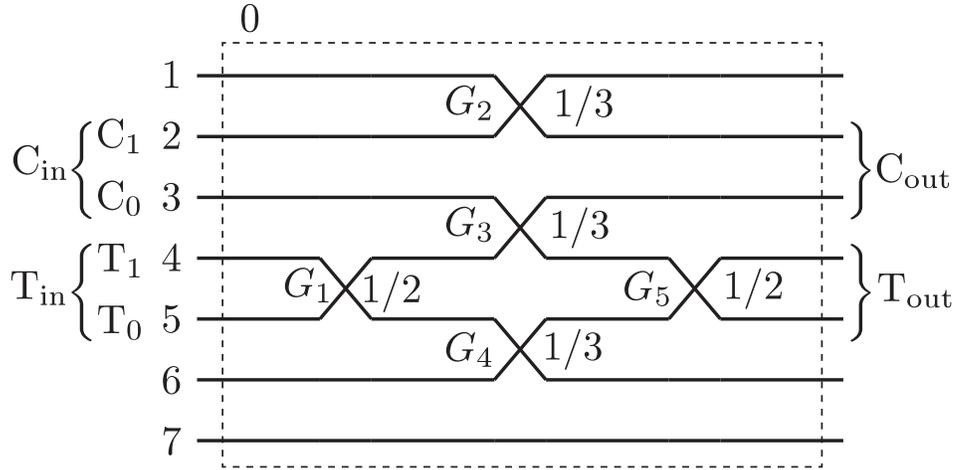}
\caption{Schematic of CNOT gate showing the five elementary beam splitting operations.  The auxiliary state is labelled as mode 7, while the bus mode (dashed box) couples to all of the other modes.  The adiabatic passage connections to and from the auxiliary mode are not shown.  The input modes corresponding to the control qubit are modes 2 and 3 (C$_1$ and C$_0$ respectively), whilst the input modes correspond to the target qubit are modes 4 and 5 (T$_1$ and T$_0$ respectively).  The crossing modes indicate adiabatic beamsplitter operations and are labelled $G_1$ to $G_5$.  The reflectivity of the gates are indicated to the right of the crossing.  Note that the figure depicts the beamsplitters as abrupt operations, with gates $G_2$ to $G_4$ performed in parallel.  Instead the adiabatic gates operate continuously and sequentially in numerical order.
}
\label{fig:CNOTSchematic}
\end{figure}

The canonical coincidence-based implementation requires six photonic modes, here encoded in the spatial modes available to the photons and shown in figure~\ref{fig:CNOTSchematic}.  Modes 1-3 correspond to the modes of the control.  Mode 1 is the vacuum state for the control, mode 2 the control in the 1 state, and mode 3 the control in the 0 state.  Mode 4 is the target 1 state, mode 5 the target 0 state and mode 6 the target vacuum state.   There are then five full gate sequences, which we denote $G_1 - G_5$. $G_1$ and $G_5$ are 50:50 beam splitters, whilst $G_2-G_4$ are 1/3:2/3 beamsplitters.  Our implementation requires eight nodes, so in addition to the six modes discussed already, there is the bus mode (denoted by mode 0) that couples all of the modes via CTAP, and an auxiliary mode (mode 7) which plays the same role as the auxiliary mode for the one photon gate.  Each gate works using the methods described above, with population adiabatic transferred from the interacting modes and auxiliary mode, via the bus mode.  In the canonical CNOT gate, the 1/3:2/3 gates are performed in parallel, however in our case, due to the shared bus and auxiliary modes, all gates must be performed sequentially.

\begin{figure}[tb!]
\centering
\includegraphics[width=0.7\textwidth]{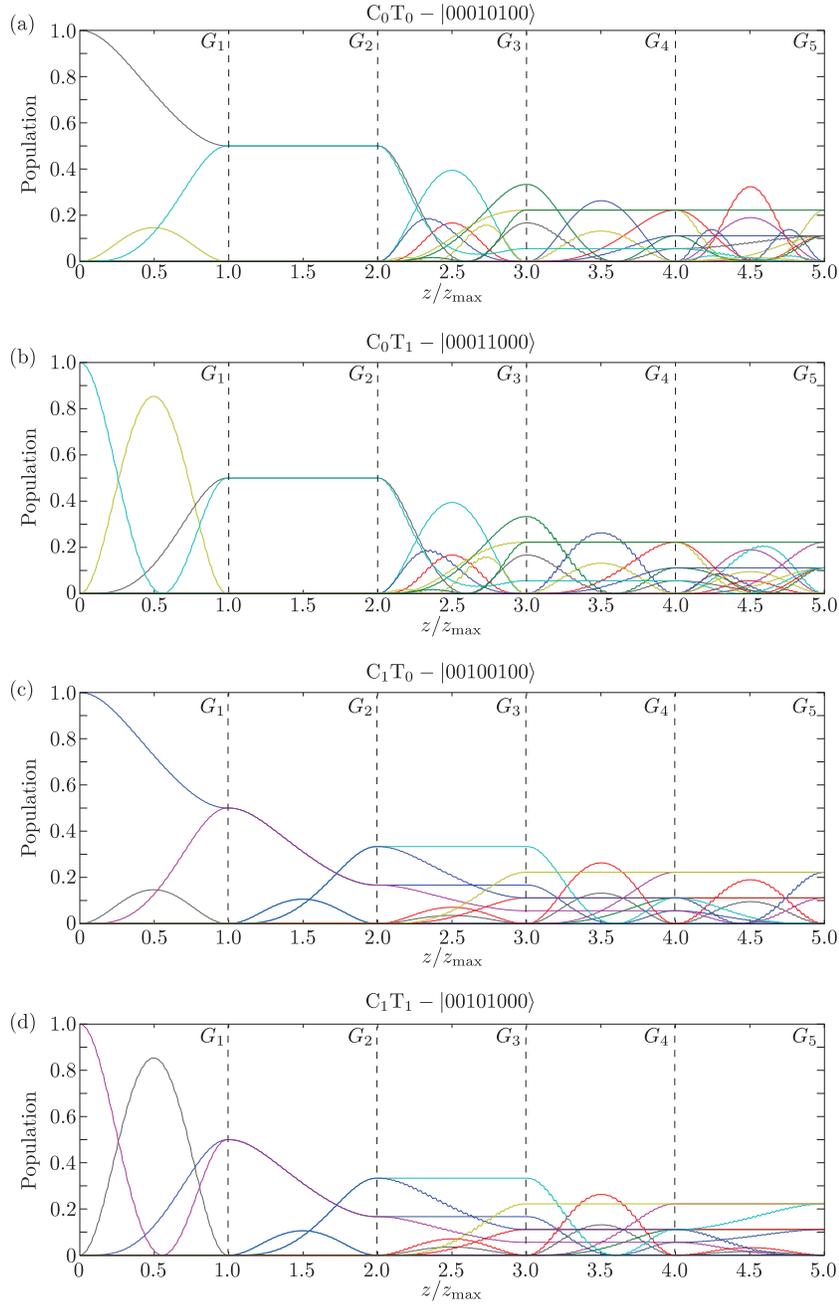}
\caption{State evolution through CTAP implementation of nondeterministic CNOT gate in the coincidence basis.  Starting states for each trace are (a) C$_0$T$_0$, (b) C$_0$T$_1$, (c) C$_1$T$_0$, and (d) C$_1$T$_1$.  Final states are listed in Table~\ref{tab:CNOTResults}, and $G_1$ to $G_5$ are the periods over which the gates are applied, as defined in text and figure~\ref{fig:CNOTSchematic}.
}
\label{fig:CNOTResults}
\end{figure}

\begin{table}[tb]
\caption{Truth table/output modes for the adiabatic CNOT gate operation.  State definitions are in the text, and designation specifies whether the output state is a failure mode or heralded success.}
\begin{center}
\begin{tabular}{|c|c|c|c|}
\hline Input state/configuration  & Output Configuration & Probability & Designation \\ \hline
\multirow{7}{*} {C$_0$T$_0$  - $\ket{00010100}$} & $\ket{00000110}$ & 1/9 & failure \\
& $\ket{00001010}$ & 1/9 & failure \\
& $\ket{00001100}$ & 1/9 & failure \\
& $\ket{00002000}$ & 2/9 & failure \\
& $\ket{00010010}$ & 1/9 & failure \\
& $\ket{00010100}$ & 1/9 & success \\
& $\ket{00020000}$ & 2/9 & failure \\ \hline
\multirow{7}{*} {C$_0$T$_1$  - $\ket{00011000}$} & $\ket{00000110}$ & 1/9 & failure \\
& $\ket{00000200}$ & 2/9 & failure \\
& $\ket{00001010}$ & 1/9 & failure \\
& $\ket{00001100}$ & 1/9 & failure \\
& $\ket{00010010}$ & 1/9 & failure \\
& $\ket{00011000}$ & 1/9 & success \\
& $\ket{00020000}$ & 2/9 & failure \\ \hline
\multirow{7}{*} {C$_1$T$_0$  - $\ket{00100100}$} & $\ket{00100010}$ & 1/9 & failure \\
& $\ket{00101000}$ & 1/9 & success \\
& $\ket{00110000}$ & 1/9 & failure \\
& $\ket{01000010}$ & 2/9 & failure \\
& $\ket{01001000}$ & 2/9 & failure \\
& $\ket{01010000}$ & 2/9 & failure \\ \hline
\multirow{7}{*} {C$_1$T$_1$  - $\ket{00101000}$} & $\ket{00100010}$ & 1/9 & failure \\
& $\ket{00100100}$ & 1/9 & success \\
& $\ket{00110000}$ & 1/9 & failure \\
& $\ket{01000010}$ & 2/9 & failure \\
& $\ket{01000100}$ & 2/9 & failure \\
& $\ket{01010000}$ & 2/9 & failure \\ \hline

\end{tabular}
\end{center}
\label{tab:CNOTResults}
\end{table}%

The results of performing the full CNOT gate operation on the appropriate starting states is shown in figure~\ref{fig:CNOTResults}. The total state space for the problem with eight modes and up to two photons per mode has dimensionality 6,561.  However only 64 states actually participate in the problem, and in the adiabatic limit, only 49 of these will have non-zero population.  Nevertheless, we do not label all of the occupied modes in the evolution shown in figure~\ref{fig:CNOTResults}, instead only highlighting the starting states, with the final states given in table~\ref{tab:CNOTResults}.  The various output configurations are labelled as success or failure on the basis of whether they correspond to heralded success or failure of the non-deterministic gate.  As expected, the table shows that the adiabatic passage CNOT gate operates in the same way as a conventional, coincidence-basis CNOT gate \cite{bib:RLB2002}, with the correct state appearing with probability 1/9.  Exactly the same heralding steps as are utilised in conventional linear optical implementations of the CNOT gate can be used in the adiabatic version.

\section{Conclusions}
We have shown several designs based on adiabatic long-range couplers that can be useful in the distribution of power within integrated photonic circuits and 
have also demonstrated how this concept can be extended to quantum information processing specifically in the form of Hadamard and NOT gates, 1/3:2/3 splitters and describe how to perform an arbitrary X-Z rotation on a photonic qubit.  We extend this to demonstrate a two-photon Hong-Ou-Mandel effect, and show the design of a complete non-deterministic linear optical CNOT gate.  The feasibility to realise these designs in a planar CMOS compatible platform is very attractive in regards to large scale integration, fabrication accuracy and circuit complexity.  While this work focusses on thin, shallow ridge SOI waveguides, it can be applied to other high index contrast materials that exhibit an accessible long-range bus mode.  This technique may be useful in realising more general Morris-Shore type devices where large scale integration is of interest.  As discussed, this scalability can require an increase in the overall device length to accommodate additional waveguides, as the overlap between each waveguide to a single bus mode decreases with increased slab width.  The adiabatic nature of these devices results in robust and repeatable signal transfer which is insensitive to variations in the device length.

\ack
Anthony Hope, Thach Nguyen and Arnan Mitchell acknowledge the support of the Australian Research Council (ARC) Centre of Excellence Funding (CE110001018), Anthony Hope acknowledges Robert and Josephine Shanks Scholarship, Thach Nguyen acknowledges support from the ARC APD fellowship (DP1096153), Andrew Greentree acknowledges the ARC for financial support (DP130104381) and useful conversations with Andon Rangelov and Alberto Peruzzo.

\end{document}